\documentclass[12pt,showpacs,preprintnumbers,superscriptaddress,amsmath,amssymb,nofootinbib]{revtex4}

\usepackage{graphicx}
\usepackage{dcolumn}
\usepackage{bm}
\usepackage{amssymb}
\usepackage{amsmath}
\usepackage{epsfig}    
\usepackage{color}
\usepackage{slashed}

\def\be{\begin{equation}}
\def\ee{\end{equation}}
\newcommand{\bea}{\begin{eqnarray}}
\newcommand{\eea}{\end{eqnarray}}
\newcommand{\nn}{\nonumber}

\numberwithin{equation}{section}

\begin{document} 
\title{Two Loop Radiative Seesaw Model with Inert Triplet Scalar Field }
\preprint{KIAS-P13013}

\author{Yuji Kajiyama}
\email{kajiyama-yuuji@akita-pref.ed.jp}
\affiliation{Akita Highschool, Tegata-Nakadai 1, Akita, 010-0851, Japan}
\author{Hiroshi Okada}
\email{hokada@kias.re.kr}
\affiliation{School of Physics, KIAS, Seoul 130-722, Korea}
\author{Kei Yagyu}
\email{keiyagyu@ncu.edu.tw}
\affiliation{Department of Physics and Center for Mathematics and Theoretical Physics,
National Central University, Chungli, Taiwan 32001, ROC}

\begin{abstract}
We propose a radiative seesaw model with an inert triplet scalar field 
in which Majorana neutrino masses are generated at the two loop level.
There are fermionic or bosonic dark matter candidates in the model. 
We find that each candidate can satisfy the WMAP data when its mass
is taken to be around the half of the mass of the standard model like Higgs boson.
We also discuss phenomenology of the inert triplet scalar bosons, especially focusing on 
the doubly-charged scalar bosons 
at Large Hadron Collider in parameter regions constrained by the electroweak precision data and WMAP data.
We study how we can distinguish our model 
from the minimal Higgs triplet model.
\end{abstract}
\maketitle
\newpage

\section{Introduction}

Recent several experiments require the serious modifications of the standard model (SM) in spite of the great success.
For an example, the SM Higgs boson search in the diphoton mode $h\to \gamma\gamma$ at Large Hadron Collider (LHC) is shown that
its signal strength is $1.65\pm0.24$ at ATLAS~\cite{ATLAS_Higgs,Moriond_gamgam} and $1.6\pm0.4$ at CMS~\cite{CMS_Higgs}.
For another example, the existence of non-baryonic dark matters (DMs), which cannot be included in SM, dominates about 23\% from the CMB observation by WMAP \cite{wmap}.
The fact is strongly supported  by the cosmological observations such as the rotation curves of the galaxy~\cite{Begeman:1991iy} and the gravitational lensing~\cite{Massey:2007wb} in our universe.
In recent years, direct detection experiments of DM; XENON100 \cite{xenon100}, CRESSTII \cite{cresst}, CoGeNT \cite{cogent} and DAMA \cite{dama}, show the scattering events with nuclei.
XENON100 has not shown a result of DM signal but shown an upper bound with the minimal bound around 100 GeV.
On the other hand, CoGeNT, DAMA and CRESSTII have reported the observations which can be interpreted as DM signals that favor a light DM with several GeV mass and rather large cross section.
As far as we consider these experiments, the mass scale of DM should be ${\cal O}$(1-100) GeV.

In order to explain the excess in the $h\to \gamma\gamma$ channel, 
a modified diphoton event has been discussed in a lot of paper. 
If a model contains charged new particles which couple to the SM-like Higgs boson such as charged Higgs bosons, 
the decay rate of $h\to \gamma\gamma$ 
can be enhanced due to loop effects of these charged particles. 
However, in a model with only one pair of singly-charged scalar bosons such as two Higgs doublet models, 
it is difficult to predict around 60\% enhancement of the decay rate unless the mass of the charged scalar bosons
is taken to be smaller than about 100 GeV\footnote{If the decay rate of the Higgs to $b\bar{b}$ mode 
is sufficiently suppressed compared to the SM value, the branching ratio of the Higgs to diphoton mode can be enhanced without changing 
its decay rate. In that case, the branching fraction of the Higgs to $WW^*$ and $ZZ^*$ modes 
are also enhanced~\cite{Chiang_Yagyu_2HDM}.}. 
The minimal Higgs triplet model (HTM) motivated from the type II seesaw mechanism \cite{typeII}, 
introducing the isospin triplet scalar field $\Delta$, can easily explain the diphoton anomaly.
Moreover, if $\Delta$ can be inert scalar, then its neutral component can be a promising DM candidate that can strongly correlate with neutrinos. This is like a radiative seesaw models
\cite{Ma:2006km, Aoki:2013gzs,Krauss:2002px,Aoki:2008av,Schmidt:2012yg, Bouchand:2012dx, Ma:2012ez, Aoki:2011he, Ahn:2012cg, Farzan:2012sa, Bonnet:2012kz, Kumericki:2012bf,
Kumericki:2012bh, Ma:2012if, Gil:2012ya, Okada:2012np, Hehn:2012kz, Dev:2012sg, Kajiyama:2012xg, Okada:2012sp, Aoki:2010ib, Kanemura:2011vm, Lindner:2011it,
Kanemura:2011mw1,
Kanemura:2011mw2,Kanemura_Sugiyama, Gu:2007ug, Gu:2008zf, Gustafsson, two-triplet}.
Hence the theory can be realized at TeV scale, so be well-tested at current experiments like LHC.

In this paper,
we propose a two-loop induced neutrino model with gauged $B-L$ symmetry that is an extension of the HTM. 
In the bosonic sector, three scalar fields are introduced in addition to the SM particles; $\Delta$ and $\eta$ 
are an $SU(2)_L$ triplet and doublet fields, respectively which do not have vacuum expectation values (VEVs), 
and $\chi$ is $SU(2)_L$ singlet which acquires a VEV after the spontaneous $B-L$ symmetry breaking
\footnote{There exists a $B-L$ gauge boson, however we neglect through our analysis because it can decouple to the other particles from the LEP II \cite{Carena:2004xs}.}. 
In the fermionic sector, we introduce an exotic vector-like lepton with $SU(2)_L$ doublet \cite{Arina:2012aj}, and three right-handed neutrino with $SU(2)_L$ singlet,
both of which can contribute to the radiative neutrino mass. Due to the abundant extra fields, we have several DM candidates.
Also we can discuss the testability of the Higgs sector, especially, the doubly-charged scalar boson, in which we could have a discrimination to the HTM, since it couples to the exotic lepton.
As a notice, such a complicated model can be realized within non-Abelian symmetries. So we briefly show how to realize our model in the appendix.

This paper is organized as follows.
In Sec.~II, we show our model building including the Higgs potential, stationary condition, neutrino mass, and lepton flavor violation (LFV). 
In Sec.~III, we analyze DM phenomenologies. In Sec.~IV, we analyze Higgs phenomenology including electroweak precision observables, 
signatures of the doubly-charged scalar boson at LHC. 
We summarize and conclude in Sec.~V. In appendices, some result of detail calculations are write down; gauge boson two point functions, decay rate of the doubly-charged scalar boson,
and the assignments for each particles in non-Abelian symmetries.

\section{The Radiative Seesaw Model}
\subsection{Model setup}

\begin{table}[thbp]
\centering {\fontsize{10}{12}
\begin{tabular}{||c|c|c|c|c|c|c||c|c||}
\hline\hline ~~Particle~~ & ~~$Q$~~ & ~~$u^c$ & $ d^c $~~ & ~~$L$~~ & ~~$e^c$~~ & ~~ $N^c$~~
  & $L'$  & $L'^c$ \\\hline
$(SU(2)_L,U(1)_Y)$ & $(\bm{2},1/6)$ & $(\bm{1},-2/3)$ & $(\bm{1},1/3)$ & $(\bm{2},-1/2)$ & $(\bm{1},1)$  & $(\bm{1},0)$  & $(\bm{2},-1/2)$ & $(\bm{2},1/2)$\\\hline
$U(1)_{B-L}$ & $1/3$ & $-1/3$ & $-1/3$ & $-1$ & $1$ & $1$  & $1$ & $-1$ \\\hline
$\mathbb{Z}_2$ & $+$ & $+$ & $+$ & $+$ & $+$  & $+$  & $-$ & $-$\\
\hline
\end{tabular}%
} \caption{The particle contents and the charges for fermions. 
$L'$ and $L'^c$ are exotic leptons.}
\label{tab:b-l}
\end{table}

\begin{table}[thbp]
\centering {\fontsize{10}{12}
\begin{tabular}{||c|c|c|c|c||}
\hline\hline ~~Particle~~ & ~~$\Delta$~~ & ~~$\Phi~~ $& ~~$\eta~~ $ & $\chi $ \\\hline
$(SU(2)_L,U(1)_Y)$ & $(\bm{3},1)$ &  $(\bm{2},1/2)$ & $(\bm{2},1/2)$ & $(\bm{1},0)$ 
\\\hline
$U(1)_{B-L}$ &  $0$ & $0$ & $0$ & $-2$  \\\hline
$\mathbb{Z}_2$ & $-$ & $+$ & $-$ & $+$  \\\hline
\end{tabular}%
} \caption{The particle contents and the charges for bosons. }
\label{tab:b-l}
\end{table}

We propose a two-loop radiative seesaw model with $U(1)_{B-L}$ gauge symmetry
which is an extended model of the minimal HTM motivated from the type II seesaw mechanism~\cite{typeII}. 
The particle contents are shown in Table~\ref{tab:b-l}. 
We add three right-handed neutrinos $N^c$, vector-like $SU(2)_L$ doublet leptons $L'$ and $L^{\prime c}$, 
an $SU(2)_L$ triplet scalar $\Delta$,
an $SU(2)_L$ doublet scalar $\eta$ and $B-L$ charged scalar $\chi$ to the SM, where
 $\eta$ and $\Delta$ do not have VEV.
The $\mathbb{Z}_2$ parity is also imposed so as to forbid the undesired terms.
As a result, the neutrino mass is obtained not through the one-loop level (just like Ma-Model ~\cite{Ma:2006km})
but through the two-loop level and the stability of DM candidates can be assured. Where we define $L'\equiv(N_{4},E_{4})$ \cite{Arina:2012aj}. 

The renormalizable Lagrangians for Yuakawa sector and Higgs potential are given by
\begin{align}
-\mathcal{L}_{\text{Yukawa}}
&=
y_{\ell}^{\alpha\beta}\Phi^\dag e^c_\alpha L_\beta + y^\gamma_{\nu}\eta^\dag N^c_\gamma L'^c + y_{\Delta}^\delta \bar L^c_\delta i\tau_2\Delta L'+y_S^{ab}\chi N^c_aN^c_b + M  L'  L'^c
+\mathrm{h.c.},\label{main-lag}\\
-\mathcal{L}_{\text{Higgs}}
&=
m_1^{2} \Phi^\dagger \Phi + m_2^{2} \eta^\dagger \eta  + m_3^{2} \chi^\dagger \chi + m_4^{2}  {\rm Tr}[\Delta^\dag\Delta]
-\mu[\Phi^Ti\tau_2\Delta^\dag\eta+{\rm h.c.}]\notag\\
&+\lambda_1 (\Phi^\dagger \Phi)^{2} + \lambda_2 
(\eta^\dagger \eta)^{2} + \lambda_3 (\Phi^\dagger \Phi)(\eta^\dagger \eta) 
+ \lambda_4 (\Phi^\dagger \eta)(\eta^\dagger \Phi)+\lambda_5/2[(\Phi^\dag\eta)^2+{\rm h.c.}]
\notag\\
&+
\lambda_6 (\chi^\dagger \chi)^{2} + \lambda_7  (\chi^\dagger \chi)
(\Phi^\dagger \Phi)  + \lambda_8  (\chi^\dagger \chi) (\eta^\dagger \eta)
+\lambda_9{\rm Det}(\Delta^\dag\Delta)
\notag\\
&
+ \lambda_{10}(  {\rm Tr}[\Delta^\dag\Delta])^2
+
a_1 {\rm Tr}[\Delta^\dag\Delta](\eta^\dag\eta)+a_2 {\rm Tr}[\Delta^\dag\tau_i\Delta](\eta^\dag\tau_i\eta)
\notag\\
&
+b_1 {\rm Tr}[\Delta^\dag\Delta](\Phi^\dag\Phi)+b_2{\rm Tr}[\Delta^\dag\tau_i\Delta](\Phi^\dag\tau_i\Phi)+
c_1 {\rm Tr}[\Delta^\dag\Delta](\chi^\dag\chi),
\end{align}
where $\alpha,\ \beta,\ \gamma,\ \delta,\ a,\ b$ are the flavor indices.
In the scalar potential, 
the couplings $\lambda_1$, $\lambda_2$, $\lambda_6$, and $\lambda_9$ have to be positive
to stabilize the potential.  
Notice here that one straightforwardly probes the term  $\Phi N^c L$, which derives neutrino mass at tree level, cannot be forbidden by any Abelian symmetries.
If we introduce non-Abelian discrete symmetries \cite{Ishimori:2012zz, Ishimori:2010au}, one finds some successful groups
such as $T_7$ \cite{Luhn:2007sy,Cao:2010mp, Cao:2011cp, Ishimori:2012sw}
and $\Delta(27)$ \cite{Luhn:2007uq, Ma:2006ip, Ma:2007wu} to forbid the term,
remaining that $\eta$ and $\Delta$ are one generation\footnote{We show in the appendix C how to realize.}.
It implies that such an extension does not affect on analyses of dark matter and Higgs phenomenology focused on the inert scalar bosons.  
So we stay to analyze our model with flavor independent way hereafter for simplicity.

The scalar fields $\Phi$, $\chi$, $\eta$ and $\Delta$ can be parameterized as 
\begin{align}
\Phi &=\left[
\begin{array}{c}
G^+\\
\frac{1}{\sqrt{2}}(\phi^0+v+iG^0)
\end{array}\right],\quad\chi=\frac{1}{\sqrt{2}}(\chi_R+v'+i\chi_I),\label{fields}\\
\eta &=\left[
\begin{array}{c}
\eta_W^+\\
\frac{1}{\sqrt{2}}(\text{Re}\eta^0+i\text{Im}\eta^0)
\end{array}\right],\quad
\Delta = \left[ 
\begin{array}{cc}
\frac{\Delta_W^+}{\sqrt{2}} & \Delta^{++} \\
\Delta_W^0 & -\frac{\Delta_W^+}{\sqrt{2}}
\end{array}\right], \text{ with }\Delta_W^0 =\frac{1}{\sqrt{2}}[\text{Re}\Delta^0+i\text{Im}\Delta^0], 
\end{align}
where $v$ is VEV of the doublet Higgs field $\Phi$ satisfying $v^2=1/(\sqrt{2}G_F)\simeq(246$ GeV$)^2$,  
$v'$ is that of the singlet Higgs field $\chi$, and subscript $W$ denotes weak eigenstates. 
In Eq.~(\ref{fields}), $G^\pm$ and $G^0$ ($\chi_I$) are (is) the Nambu-Goldstone bosons (boson)
which are (is) absorbed by the longitudinal component of $W^\pm$ and $Z$ (additional $U(1)_{B-L}$ gauge boson). 
The VEVs of $\Delta$ and $\eta$ are taken to be zero because of the assumption of the unbroken $\mathbb{Z}_2$ symmetry. 

%
In this model, the $\mathbb{Z}_2$-even fields $\Phi$ and $\chi$ cannot be mixed with the $\mathbb{Z}_2$-odd fields $\Delta$ and $\eta$, 
so that the mass matrices for the component scalar fields from $\Phi$ and $\chi$ and those from $\Delta$ and $\eta$ can be 
separately considered. 
By inserting the tadpole conditions; $m^2_1=-\lambda_1v^2-\lambda_7v'^2/2$
and $m^2_3=-\lambda_6v'^2 - \lambda_7v^2/2$ (which is exactly the same as the results in Ref.~\cite{Okada:2012np}), 
the mass matrix for the CP-even scalar bosons in the basis of ($\phi^0,\chi^0$) is calculated as
\begin{equation}
\mathcal{M}^2(\phi^0,\chi^0)
=
\left(%
\begin{array}{cc}
  2\lambda_1v^2 & \lambda_7vv'  \\
  \lambda_7vv' & 2\lambda_6v'^2 \\
\end{array}%
\right), \label{mat_even}
\end{equation}
The mass matrices of the $\mathbb{Z}_2$-odd scalar bosons are respectively  obtained in the basis of ($\Delta_W^\pm$, $\eta_W^\pm$), 
(Re$\Delta^0$, Re$\eta^0$)
and (Im$\Delta^0$, Im$\eta^0$) by 
\begin{align}
\mathcal{M}^2(\Delta_W^\pm,\eta_W^\pm) 
&=
\left[%
\begin{array}{cc}
m^{2} (\Delta_W^+) & \mu v/2  \\
 \mu v/2 &m^{2} (\eta_W^+) \\
\end{array}\right],~
\mathcal{M}^2({\rm Re}\Delta^0,{\rm Re}\eta^0) 
=
\left[%
\begin{array}{cc}
m^2( {\rm Re} \Delta^{0})  & \mu v/\sqrt2  \\
 \mu v/\sqrt2 & m^{2} ({\rm Re}\eta^{0})  \\
\end{array}%
\right],\notag\\
\mathcal{M}^2({\rm Im}\Delta^0,{\rm Im}\eta^0) 
&=
\left[%
\begin{array}{cc}
 m^{2} ( {\rm Im} \Delta^{0})  & \mu v/\sqrt2  \\
 \mu v/\sqrt2 & m^{2} ({\rm Im} \eta^{0})  \\
\end{array}%
\right],
\label{mat_odd}
\end{align}
where we define as follows:
\begin{eqnarray}
m^{2} (\Delta_W^{\pm}) &=&  m_4^{2} + \frac12 c_1v'^2 +\frac12b_1v^2,\\
 m^{2} ( {\rm Re}\Delta^{0})  &=& m^{2} ( {\rm Im}\Delta^{0})=
m^{2} (\Delta_W^{\pm})+\frac{b_2}{2}v^2,\\
m^{2} (\eta_W^{\pm}) &=&  m_2^{2} + \frac12 \lambda_3 v^{2} + \frac12 \lambda_8 v'^{2}, \\ 
m^{2} ( {\rm Re} \eta^{0}) &=& m_2^{2} + \frac12 \lambda_8 v'^{2}
 + \frac12 (\lambda_3 + \lambda_4 + \lambda_5) v^{2}, \\ 
 m^{2} (  {\rm Im} \eta^{0})  &=& m_2^{2} + \frac12 \lambda_8 v'^{2}
 + \frac12 (\lambda_3 + \lambda_4 -\lambda_5) v^{2}. 
\end{eqnarray}
The mass eigenstates for the $\mathbb{Z}_2$-even Higgs bosons and those 
for the $\mathbb{Z}_2$-odd scalar bosons can be defined by introducing the mixing angles $\alpha$, $\beta$, $\gamma$ and $\delta$ as
\begin{align}
&\left(
\begin{array}{c}
\phi^0 \\
\chi_R^0 
\end{array}
\right)=R(\alpha) \left(
\begin{array}{c}
h \\
H
\end{array}
\right),~\left(
\begin{array}{c}
\Delta_W^\pm \\
\eta_W^\pm 
\end{array}
\right)=R(\beta) \left(
\begin{array}{c}
\Delta^\pm \\
\eta^\pm 
\end{array}
\right),\notag\\ 
&\left(
\begin{array}{c}
\text{Re}\Delta^0 \\
\text{Re}\eta^0
\end{array}
\right)=R(\gamma) \left(
\begin{array}{c}
\Delta_R^0 \\
\eta_R^0 
\end{array}
\right),~
\left(
\begin{array}{c}
\text{Im}\Delta^0 \\
\text{Im}\eta^0
\end{array}
\right)=R(\delta) \left(
\begin{array}{c}
\Delta_I^0 \\
\eta_I^0 
\end{array}
\right),\notag\\
&\text{with}~~R(\theta) = \left(
\begin{array}{cc} 
\cos\theta & -\sin\theta \\
\sin\theta & \cos\theta
\end{array}\right), 
\end{align}
where $h$ can be regarded as the SM-like Higgs boson. 
These mixing angles are expressed as 
\begin{align}
\tan2\alpha &= \frac{\lambda_7 vv'}{\lambda_1v^2-\lambda_6v^{\prime 2}},\quad 
\tan2\beta = \frac{\mu v}{m^{2} (\Delta_W^{\pm})-m^{2} (\eta_W^{\pm})},\notag\\
\tan2\gamma &= \frac{\sqrt{2}\mu v}{m^2(\text{Re}\Delta^0)-m^2(\text{Re}\eta^0)},\quad 
\tan2\delta = \frac{\sqrt{2}\mu v}{m^2(\text{Im}\Delta^0)-m^2(\text{Im}\eta^0)}. \label{angles}
\end{align}
The mass eigenvalues are calculated by using the mixing angles given in Eq.~(\ref{angles}) 
and the mass matrices given in Eqs.~(\ref{mat_even}) and (\ref{mat_odd}) as
\begin{align}
R^T(\alpha)\mathcal{M}^2(\phi^0,\chi^0)R(\alpha)=
\left(
\begin{array}{cc} 
m_h^2 & 0 \\
0 & m_H^2
\end{array}\right),~
R^T(\beta)\mathcal{M}^2(\Delta_W^\pm,\eta_W^\pm)R(\beta)=
\left(
\begin{array}{cc} 
m_{\Delta^+}^2 & 0 \\
0 & m_{\eta^+}^2
\end{array}\right),\notag\\
R^T(\gamma)\mathcal{M}^2(\Delta_R^0,\eta_R^0)R(\gamma)=
\left(
\begin{array}{cc} 
m_{\Delta_R}^2 & 0 \\
0 & m_{\eta_R}^2
\end{array}\right),~R^T(\delta)\mathcal{M}^2(\Delta_I^0,\eta_I^0)R(\delta)=
\left(
\begin{array}{cc} 
m_{\Delta_I}^2 & 0 \\
0 & m_{\eta_I}^2
\end{array}\right). 
\end{align}
The doubly-charged triplet scalar bosons do not mix with others, and those masses are given by 
\begin{eqnarray}
m^2_{\Delta^{++}} =  m^{2} (\Delta_W^{\pm})-\frac{1}{2}b_2v^2. 
\end{eqnarray}

We note that there is a characteristic relations for the mass spectrum among the 
triplet-like scalar bosons $\Delta^{\pm\pm}$, $\Delta^\pm$, $\Delta_R^0$ and $\Delta_I^0$ in 
the limit of $\mu\to 0$ as
\begin{align}
&m_{\Delta^{++}}^2-m_{\Delta^+}^2=m_{\Delta^+}^2-m_{\Delta^0}^2, \label{mass1}\\
&m_{\Delta^0}^2=m_{\Delta_R}^2=m_{\Delta_I}^2.
\end{align}
Therefore, two of four mass parameters for the triplet-like scalar bosons are determined by using above two equations. 
The same relations also appear in HTM when the lepton number violating coupling constant is taken to be zero~\cite{Mass_relations}.

\subsection{Neutrino mass matrix}
\begin{figure}[t]
\begin{center}
 \includegraphics[width = 100mm]{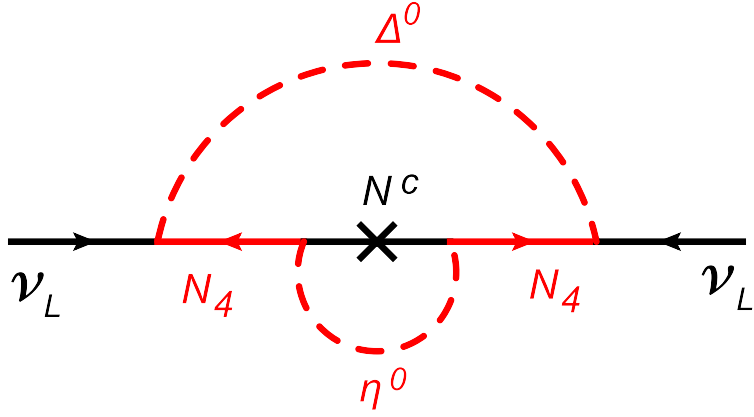}
   \caption{Neutrino mass generation via two-loop radiative seesaw. 
The particles indicated by a red font have the opposite $\mathbb{Z}_2$ charge to those by a black font. }
   \label{three-neut}
\end{center}
\end{figure}

The active neutrino mass matrix~(depicted in Fig.\ref{three-neut}) through two loop contribution is given by

\begin{eqnarray}
\left(M_\nu\right)_{\alpha\beta}&=& 
\frac{(y_\Delta)_{\alpha } (y_\nu)^2_{k}(M_{N_k}) M^2 (y_\Delta)_{\beta}}{4}
\int\frac{d^4p}{(2\pi)^4}\int\frac{d^4q}{(2\pi)^4}\frac{1}{(p^2-M^2)^2}\frac{1}{{(p+q)^2-M^2_{N_k}}}\notag\\
&\times& \left( \frac{\sin^2\gamma}{p^2-m^2_{\eta_R}} +  \frac{\cos^2\gamma}{p^2-m^2_{\Delta_R}} -\frac{\sin^2\delta}{p^2-m^2_{\eta_I}}
 - \frac{\cos^2\delta}{p^2-m^2_{\Delta_I}}\right)\notag\\
&\times& \left( \frac{\cos^2\gamma}{q^2-m^2_{\eta_R}} +  \frac{\sin^2\gamma}{q^2-m^2_{\Delta_R}} -\frac{\cos^2\delta}{q^2-m^2_{\eta_I}}
 - \frac{\sin^2\delta}{q^2-m^2_{\Delta_I}}\right),
\end{eqnarray}
where
\begin{align}
&\int\frac{d^4p}{(2\pi)^4}\int\frac{d^4q}{(2\pi)^4}\frac{1}{(p^2-M^2)^2}\frac{1}{(p+q)^2-M^2_{N_k}} 
\frac{1}{p^2-m^2_{a}} \frac{1}{q^2-m^2_{b}}\notag\\
&=\frac{1}{(4\pi)^4}\int^1_0 dx \int^{1-x}_0 dz\int^1_0 d\rho \frac{x \rho}{(z^2-z)^2}\frac{1}{(1-\rho)M_{N_a}-\rho q^2},\\
q^2&=\frac{x}{z^2-z}M^2+\frac{1-x-z}{z^2-z}m^2_{a}+\frac{1}{z-1}m^2_b.
\end{align}
As can be seen in the above equation, we can reproduce observed neutrino masses $\sim{\cal O}$(0.1) eV  and the mixing data because of many parameters.

\subsection{Lepton Flavor Violation}
\begin{figure}[t]
\begin{center}
 \includegraphics[scale=1.0]{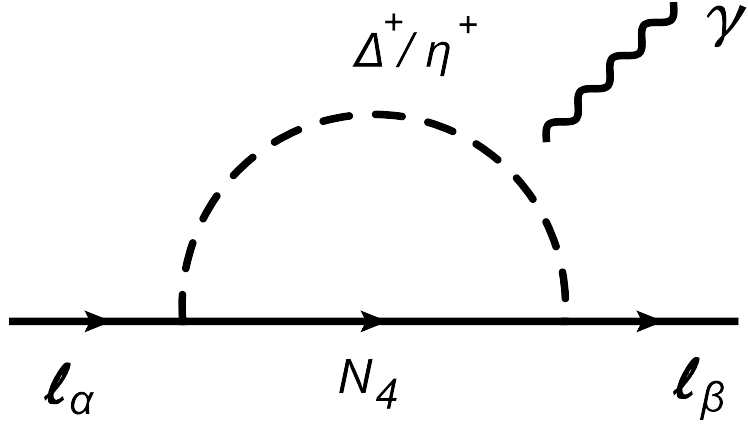}
   \caption{Lepton flavor violation.}
   \label{fig:lfv}
\end{center}
\end{figure}

We investigate the LFV process $\ell_{\alpha} \to \ell_{\beta}\gamma~(\ell_\alpha,~\ell_\beta=e,\mu,\tau)$
as shown in Fig. \ref{fig:lfv}.
The experimental upper bounds of the branching ratios are
${\cal B}\left(\mu\to e\gamma\right)\leq2.4\times 10^{-12}$~\cite{Adam:2011ch},
${\cal B}\left(\tau\to\mu\gamma\right)\leq4.4\times10^{-8}$ and
${\cal B}\left(\tau\to e\gamma\right)\leq3.3\times10^{-8}$~\cite{Nakamura:2010zzi}. 
The branching ratios of the processes
$\ell_{\alpha}\rightarrow\ell_{\beta}\gamma$ are calculated as
 \begin{equation}
 {\cal B} (\ell_{\alpha}\rightarrow\ell_{\beta}\gamma)
=
\frac{3\alpha_{\mathrm{em}}}{64\pi G^2_{F}}
\left| y^{\dag}_{\Delta_\alpha} y_{\Delta_\beta}
\left[ F_2\left(\frac{M^2}{m_{\eta^+}^2}\right) \frac{\cos^2\beta}{m^{2}_{\eta^+}}
+ F_2\left(\frac{M^2}{m_{\Delta^+}^2}\right)\frac{\sin^2\beta}{m^{2}_{\Delta^+}} \right]  \right|^{2}
{\cal B}(\ell_{\alpha}\rightarrow\ell_{\beta}\overline{\nu_{\beta}}\nu_{\alpha}),
 \label{LFV1}
 \end{equation}
where $\alpha_{\mathrm{em}}=1/137$, ${\cal B}\left(\mu\to
e\overline{\nu_e}\nu_\mu\right)=1.0$, ${\cal B}\left(\tau\to
e\overline{\nu_e}\nu_\tau\right)=0.178$,
${\cal B}\left(\tau\to\mu\overline{\nu_\mu}\nu_\tau\right)=0.174$, 
$G_{F}$ is the Fermi constant and 
the loop function $F_{2}(x)$ is given by
\begin{eqnarray}
F_{2}(x)=\frac{1-6x+3x^{2}+2x^{3}-6x^{2}\ln{x}}{6(1-x)^{4}}.
\end{eqnarray}
The $\mu\to e\gamma$ process gives the most stringent constraint. 
The mixing angle $\sin\beta$ directly does not contribute to the neutrino mass, however,  
$\sin2\beta$ should not be zero to retain $\mu v\neq 0$ in Eq.~(\ref{angles}).

\section{Dark Matters}
We discuss DM candidates in this section. We have six DM candidates in general; that is, $\eta_{R(I)}$,  $\Delta_{R(I)}$,
the lightest one of $N^c$, and $N_4$. However $N_4$ cannot be the candidate because $N_4 \bar N_4$ annihilation via $Z$ boson
gives too large cross section to obtain the observed relic density $\Omega h^2\simeq$ 0.11 \cite{wmap} as well as too large
scattering cross section in the direct detection search \cite{xenon100}
\footnote{ In Ref. \cite{Arina:2012aj, Joglekar:2012vc}, $N_4$ is considered as a DM. }. Moreover, we restrict ourselves that
only the real part of $\eta_{R(I)}$ and $\Delta_{R(I)}$ are considered as a DM candidates,
since the DM property of the imaginary part is more or less the same as the real one. 
Hence we analyze three DM candidates; $\eta_{R}$,  $\Delta_{R}$, the lightest one of $N^c$, below.
Hereafter we symbolize the DM mass as $m_{\rm DM}$.
\begin{figure}[t]
\begin{center}
 \includegraphics[scale=0.7]{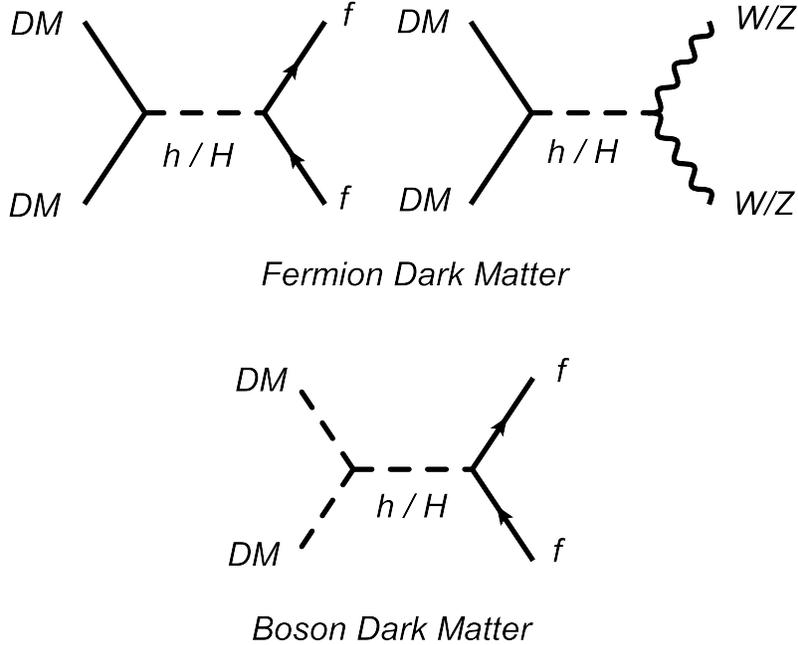}
   \caption{Dominant annihilation channel of DM. The left figure is for the fermionic DM; $N^c$, 
and the right one is for the bosonic ones; $\eta_R$ and $\Delta_R$.}
   \label{fig:ann}
\end{center}
\end{figure}

\subsection{Fermionic Dark Matter}
We discuss a fermionic DM candidate $N^c$, assuming the following mass hierarchy $M_1<M_2<M_3$ for the right-handed neutrinos $N^c_i$.
Notice here that the DM mass range be less than $M=$ 100 GeV to avoid the too short lifetime of DM.

{\it\underline{WMAP}}: At first, we analyze the DM relic density from WMAP.
We have only the s-channel process via the Higgs bosons $h/H$ as shown in the upper panel of Fig.~\ref{fig:ann}. The effective cross section to
$f\overline{f}/W^+W^-/2Z^0$ is given as
 \begin{align}
\sigma^{N^c}_{\mathrm{eff}} v_{\rm rel}
&\simeq
\frac{m_{\rm DM}^2\sin^2\alpha\cos^2\alpha}{8\pi v'^2 v^2}
\left|\frac{1}{4m_{\rm DM}^2-m_{h}^2+im_{h}\Gamma_h}
-\frac{1}{4m_{\rm DM}^2-m_{H}^2+im_{H}\Gamma_H}\right|^2v^2_{\rm rel}\notag\\
&\times
\sum_{i=f,V}\left(1-\frac{m_i^2}{m_{\rm DM}^2}\right)^{1/2}
\left[3m_i^2 m_{\rm DM}^2 \left(1-\frac{m_i^2}{m_{\rm DM}^2}\right)\delta_{i,f}
+m_i^4 \left(\frac32 - 2\frac{m_\text{DM}^2}{m_{i}^2} + 2\frac{m_\text{DM}^4}{m_{i}^4}\right)\delta_{i,V}\right]\notag\\
&\simeq
\frac{m_{\rm DM}^2\sin^2\alpha\cos^2\alpha}{8\pi v'^2 v^2}
\left|\frac{1}{4m_{\rm DM}^2-m_{h}^2+im_{h}\Gamma_h}\right|^2v^2_{\rm rel}
\sum_{i=b,V}\left(1-\frac{m_i^2}{m_{\rm DM}^2}\right)^{1/2}
\notag\\
&\times
\left[3m_i^2 m_{\rm DM}^2 \left(1-\frac{m_i^2}{m_{\rm DM}^2}\right)\delta_{i,b}
+m_i^4 \left(\frac32 - 2\frac{m_\text{DM}^2}{m_{i}^2} + 2\frac{m_\text{DM}^4}{m_{i}^4}\right)\delta_{i,V}\right],
\label{eq:ann}
 \end{align}
where $V=W^\pm,Z^0$ and we neglected the light quarks
contributions and put only bottom quark contribution; that is, $f\simeq b$, 
and $v_{\rm rel}$; $v_{\rm rel}\simeq{\cal O} (0.2)$ \cite{Cirelli:2008pk}, is the relative velocity of incoming DM. 
Here we neglect the contribution of $H$, assuming the mass is enough heavy.
The SM-like Higgs mass is fixed to
$m_h=125~\mathrm{GeV}$. Notice here that the channel of $VV$ is opened only if $m_V\le m_{\rm DM}$.  In $m_b\le m_h/2\le m_{\rm DM}$,
the total decay width of $h$ is $\Gamma_h=4.1\times10^{-3}~\mathrm{GeV}$~\cite{higgsdecay}. 
Moreover, in $ m_{\rm DM}\le m_h/2$, the channel $h\rightarrow 2{\rm DM}$ is also added and given by
\begin{equation}
\Gamma_h(h\to2{\rm DM})\simeq
\frac{m_h}{8\pi}\left(\frac{ m_{\rm DM}\sin\alpha}{v'}\right)^2\left[1-4\left(\frac{m_{\rm DM}}{m_h}\right)^2 \right]^{3/2},
\end{equation}
which is known as an invisible decay and recently reported by the LHC experiment 
that the branching ratio ${\cal B}_{inv}$ is excluded to the region $0.4\le{\cal B}_{inv}$ \cite{Giardino:2012dp}.

{\it\underline{Direct Detections}}: Let us move on to the discussion of direct detections.
Our DM interacts with quarks via Higgs exchange. 
Thus it is possible to explore DM in direct detection experiments
like XENON100 \cite{xenon100}. 
The Spin Independent (SI) elastic cross section $\sigma_{\mathrm{SI}}$
with nucleon $N$ is given by
\begin{equation}
\sigma_{\mathrm{SI}}^N\simeq\frac{\mu_{\mathrm{DM}}^2}{ m_h^2\pi}
\left(\frac{\left(y_S\right)_{11}m_N\sin\alpha\cos\alpha}
{\sqrt{2}v}\sum_{q}f_q^N\right)^2,
\label{eq:dd}
\end{equation}
where $\mu_{\mathrm{DM}}=\left(m_{\rm DM}^{-1}+m_N^{-1}\right)^{-1}$ is the
DM-nucleon reduced mass and the heavy Higgs contribution is neglected. 
The parameters $f_q^N$ which imply the contribution of each quark to
nucleon mass are calculated by the lattice
simulation~\cite{Corsetti:2000yq, Ohki:2008ff} as
\begin{align}
&f_u^p=0.023,\quad
f_d^p=0.032,\quad
f_s^p=0.020,\\
&f_u^n=0.017,\quad
f_d^n=0.041,\quad
f_s^n=0.020,
\end{align}
for the light quarks and $f_Q^N=2/27\left(1-\sum_{q\leq 3}f_q^N\right)$ for
the heavy quarks $Q$ where $q\leq3$ implies the summation of the light
quarks. The recent another calculation is performed in Ref.~\cite{Alarcon:2011zs}.

Considering all the above constraints, we find that the allowed region is sharp at around $m_h/2$.

\subsection{Bosonic Dark Matters}
We discuss bosonic DM candidates $\eta_R$ and $\Delta_R$ \cite{triplet-dm}.
Notice here that the DM mass range be less than $M_W=$ 80 GeV
to satisfy the constraint of the antiproton no excess reported by PAMELA \cite{Adriani:2010rc} as well as WMAP \footnote{If the charged vector boson channel are open, the cross section is too large to satisfy the
observed relic abundance.}.

{\it\underline{WMAP}}: At first, we analyze the DMs relic density from WMAP.
We have two annihilation modes; 
$t$ and $u$ channel of $2\eta_R/\Delta_R\to N_4(E_4) \to 2\nu(\ell\bar\ell)$
and $s$-channel of  $2\eta_R/\Delta_R \to h/H \to f\bar f$ \footnote{We neglect the $H$ contribution.}.
However since the $t$ and $u$ channel does not have $s$ wave contribution,
the dominant cross section is given only through the $s$ channel \cite{Kajiyama:2012xg}
as shown in the lower diagram of Fig.~\ref{fig:ann}.
 The effective cross section to
$f\overline{f}$ is given as
\begin{eqnarray}
\sigma^{\eta_R/\Delta_R}_{\mathrm{eff}} v_{\rm rel}
&\simeq&
\frac{3m^2_b}{4\pi v^2}
\left|\frac{\lambda^{\eta_R/\Delta_R}_h\cos\alpha}{4m_{\rm DM}^2-m_{h}^2+im_{h}\Gamma_h}\right|^2
\left(1-\frac{m_b^2}{m_{\rm DM}^2}\right)^{3/2},
\label{b-sigma}\\
\lambda^{\eta_R}_h&\equiv& 
\cos\alpha\left[\cos^2\gamma(\lambda_3+\lambda_4+\lambda_5)+\sin^2\gamma(b_1+b_2)-\sqrt2\mu\sin\gamma\cos\gamma\right]v\notag\\
&&-\sin\alpha(\cos^2\gamma\lambda_8+\sin^2\gamma c_1)v',\\
\lambda^{\Delta_R}_h&\equiv& 
\cos\alpha\left[\sin^2\gamma(\lambda_3+\lambda_4+\lambda_5)+\cos^2\gamma(b_1+b_2)+\sqrt2\mu\sin\gamma\cos\gamma\right]v\notag\\
&&-\sin\alpha(\sin^2\gamma\lambda_8+\cos^2\gamma c_1)v'.
\end{eqnarray}

When our DMs are less than $m_h/2$, the invisible decay width is give by \cite{Kajiyama:2012xg}
\begin{eqnarray}
\Gamma^{\eta_R/\Delta_R}(h\rightarrow 2{\rm DM})\simeq\frac{(\lambda^{\eta_R/\Delta_R}_h)^2}{16\pi m_h}\sqrt{1-4m_{\rm DM}^2/m_h^2} .
\end{eqnarray}

{\it\underline{Direct Detections}}:
In the direct detection, the spin independent elastic cross section $\sigma_{SI}$ with nucleon $N$ is given by 
\be
\sigma^{\eta_R/\Delta_R}_{SI}=\frac{\mu_{\mathrm{DM}}^2}{\pi}\frac{m_N^2}{m_{\chi}^2v^2}
\left(\frac{\lambda^{\eta_R/\Delta_R}_h\cos\alpha}{m_h^2}\right)^2
\left(\sum_qf_q^p\right)^2,
\ee 
where $\mu_{\rm DM}$ and $f_q^N$ has been defined in the fermionic part.
We here comment on the scenario based on the neutral component of 
the inert triplet field being the DM candidate, which has been discussed in Ref. [63]. 
Such a scenario is severely constrained by the direct detection experiments because of 
the Z boson exchanging contribution if the CP-even scalar boson and the CP-odd scalar boson from the triplet field are degenerate in mass.  
According to Ref. [63], the mass of the DM candidate has to be around 2.8 TeV 
in the model with Y=1 inert triplet field in order to satisfy both the WMAP data and the 
direct search data. 
However, this result cannot be applied to our model, because we can take a mass splitting 
between $\Delta_R^0$ and $\Delta_I^0$ due to the mixing between $\eta$ and $\Delta$. 
Thus, we can avoid the Z boson exchanging contribution to the direct search experiments.

Considering all the above constraints, we find that the allowed region is the same as the fermionic case; that is, around $m_h/2$. 
As a result, we assume to analyze that DM mass (especially $\Delta^0$) is $m_h/2$ in the next section.

\section{Higgs phenomenology}

\subsection{Electroweak precision observables}

We discuss the constraints of the Higgs parameters from the electroweak precision observables; $i.e.$, 
the Peskin-Takeuchi $S$, $T$ and $U$ parameters~\cite{PT}. 
If the mixing angle $\alpha$ is taken to be zero, the new physics contributions to the $S$ and $T$ parameters can be 
separated from those from the SM, so that we here assume $\alpha=0$ for simplicity. 
Then the new physics contributions to $S$ and $T$ are calculated as 
\begin{align}
S_{\text{new}}&=16\pi\left[\frac{\Sigma_T^{3Q}(m_Z^2)-\Sigma_T^{3Q}(0)}{m_Z^2}-
\frac{\Sigma_T^{33}(m_Z^2)-\Sigma_T^{33}(0)}{m_Z^2}\right],\\
T_{\text{new}}&=\frac{4\sqrt{2}G_F}{\alpha_{\text{em}}}\left[\Sigma_T^{33}(0)-\Sigma_T^{11}(0)\right], 
\end{align} 
where $\Sigma_T^{11}$, $\Sigma_T^{3Q}$ and $\Sigma_T^{33}$ are obtained by calculating the 1PI diagrams  of the 
gauge boson two point functions at the one-loop level whose analytic expressions are given in Appendix~A. 
The experimental bound for the deviations in the $S$ and $T$ parameters from the 
SM prediction with the Higgs boson mass to be 117 GeV are given by fixing $\Delta U=0$ as 
\begin{align}
\Delta S=0.04 \pm 0.09,\quad \Delta T = 0.07\pm 0.08, 
\end{align}
with the correlation factor to be 88\%~\cite{Nakamura:2010zzi}. 
The prediction in our model for $\Delta S$ and $\Delta T$ are calculated by $S_\text{new}-(S_\text{SM}-S_\text{SM}|_{m_h = 117~\text{GeV}})$ and 
$T_\text{new}-(T_\text{SM}-T_\text{SM}|_{m_h = 117~\text{GeV}})$, respectively, 
where $S_{\text{SM}}$ ($S_{\text{SM}}|_{m_h=117~\text{GeV}}$) is the prediction for $S$ in the SM with the SM Higgs boson mass to be $m_h$ 
(117 GeV), and $T_{\text{SM}}$ ($T_{\text{SM}}|_{m_h=117~\text{GeV}}$) is those corresponding $T$ parameter. 
We note that the exotic lepton $L'$ does not contribute to the $S$ and $T$ parameters, because 
both the masses of the component lepton fields ($N_4$ and $E_4$) 
are given by $M$, and this field is introduced as the vector-like, so that 
both the custodial symmetry and the chiral symmetry are not broken by this field. 

\begin{figure}[t]
\begin{center}
\includegraphics[width=120mm]{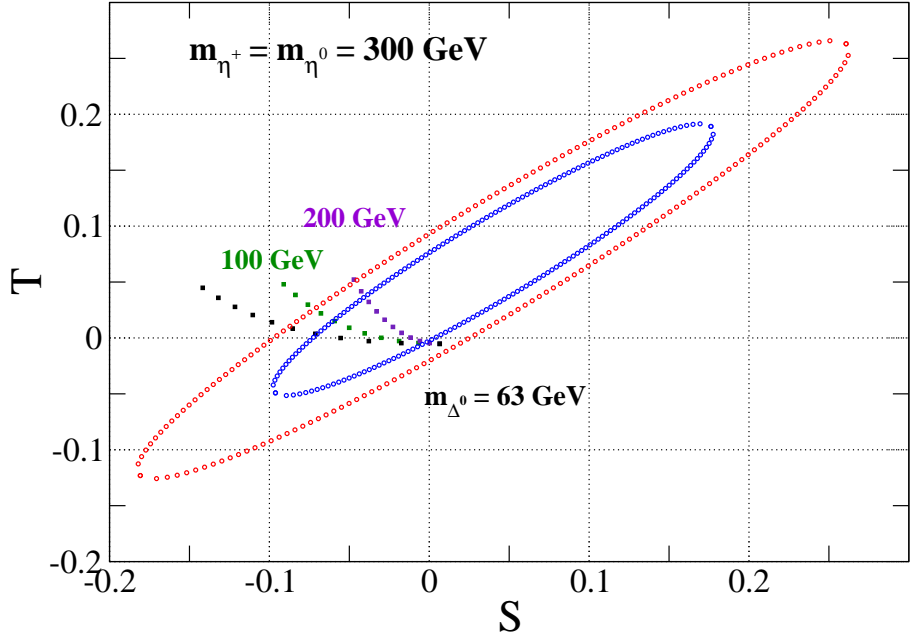}
\caption{Prediction of the $S$ and $T$ parameters for the case with $m_{\eta^+}=m_{\eta^0}(=m_{\eta_R}=m_{\eta_I})=300$ GeV. 
All the mixing angles ($\alpha$, $\beta$, $\gamma$ and $\delta$) are taken to be zero in this plot. 
Each dotted curve is shown the results in the cases with $m_{\Delta^0}$ to be $63$ GeV, 100 GeV and 200 GeV, 
where the interval of each dot indicates 
the increment of $\Delta m(=m_{\Delta^+}-m_{\Delta^0})$ to be 3 GeV from right to left. 
The regions within the blue (red) ellipse are allowed at the 68\% (95\%) confidence level by the electroweak precision data.
}
\label{ST}
\end{center}
\end{figure}

In Fig.~\ref{ST}, $S$-$T$ plot is shown in the case where $m_{\eta^+}=m_{\eta_R}=m_{\eta_I}=300$ GeV, and 
all the mixing angles are taken to be zero, so that the masses of the scalar bosons from the triplet field are determined 
by fixing two parameters; $i.e.$, $m_{\Delta^0}$ and $\Delta m(= m_{\Delta^+}-m_{\Delta^0})$. 
Each dotted curve shows the prediction of the $S$ and $T$ parameters 
in the cases with $m_{\Delta^0}=63$ GeV ($m_h/2$), 
100 GeV and 200 GeV.
The interval of each dot indicates 
the increment of $\Delta m(=m_{\Delta^+}-m_{\Delta^0})$ to be 3 GeV from right to left. 
The inside (outside) ellipse indicates allowed region with the 68\% (95\%) confidence level by the electroweak precision data. 
It is seen that that the $T$ parameter is getting larger values when $\Delta m$ is taken to be larger values. 
When $m_{\Delta^0}$ is taken to be 63 GeV, 100 GeV and 200 GeV, the allowed maximum value for $\Delta m$ with 95\% confidence level 
is about 
15 GeV, 23 GeV and 30 GeV, respectively.

\subsection{Signature of $\Delta^{\pm\pm}$ at LHC }

We discuss how our model can be tested at collider experiments. 
We focus on the signature from the doubly-charged scalar bosons $\Delta^{\pm\pm}$, because
appearance of $\Delta^{\pm\pm}$ is one of the striking properties of the model. 
In order to focus on the phenomenology of the triplet-like scalar bosons, 
we assume that the mixing between $\Delta$ and $\eta$ are taken to be quite small $(\beta\simeq \gamma\simeq \delta\simeq 0)$, 
and the masses of $\eta^\pm$ and $\eta^0$ are much heavier than those of the triplet-like scalar bosons. 
 
There are an indirect way and a direct way to identify existence of $\Delta^{\pm\pm}$. 
The farmer way is measuring the deviations in the event rate for the Higgs boson decay channels from the SM values.
In particular, the Higgs to diphoton mode $h\to \gamma\gamma$ is one of the most important channels for the SM Higgs boson search 
at the LHC because of the clear signature. 
The current signal strength for this channel is $1.65\pm0.24$ at the ATLAS~\cite{ATLAS_Higgs,Moriond_gamgam} and $1.6\pm0.4$ at the CMS~\cite{CMS_Higgs}.
The decay rate of $h\to \gamma\gamma$ can be modified 
by the loop effect of $\Delta^{\pm\pm}$. 
Contributions from doubly-charged scalar bosons to the $h\to \gamma\gamma$ mode have already been analyzed in 
the several papers in the HTM~\cite{hgg_HTM,Chiang_Yagyu}. 
The signal strength for $h\to\gamma\gamma$ can be larger than 1.6 in the case with 
$m_{\Delta^{\pm\pm}}$ to be smaller than about 200 GeV~\cite{Chiang_Yagyu} without contradiction with the constraints from the 
vacuum stability~\cite{VS_arhrib} and the perturbative unitarity~\cite{VS_arhrib, PV_AK}. 
The prediction of the deviation in the decay rate of $h\to \gamma\gamma$ in our model is 
almost the same as that in the HTM as long as the contributions from $\eta^\pm$ 
are neglected.

Discovery of $\Delta^{\pm\pm}$ can be a direct evidence of our model. 
We consider the case where a lighter neutral scalar boson from the triplet field ($\Delta_R$ or $\Delta_I$)
is assumed to be DM.
In that case, the neutral scalar boson should be the lightest of all the triplet-like scalar bosons; $i.e.,$
$m_{\Delta^0} \leq m_{\Delta^+}\leq m_{\Delta^{++}}$ to guarantee the stability of DM, 
and its mass is around the half of the Higgs boson mass $m_h$ to satisfy WMAP data and direct detection experiments. 
In addition, the mass differences among the triplet-like scalar bosons 
are restricted from the $S$ and $T$ parameter as shown in Fig.~\ref{ST}; $e.g.,$
$\Delta m$ is constrained to be less than 15 GeV in the case with $m_{\Delta^0}=63$ GeV. 
This implies that the upper limit for $m_{\Delta^{++}}$ is about 90.5 GeV by using the mass relation given in Eq.~(\ref{mass1}). 

A search for doubly-charged Higgs bosons has been done at LEP~\cite{H++_LEP}, Tevatron~\cite{H++_TeV} and LHC~\cite{H++_LHC}. 
All these searches have been done under the assumption where doubly-charged Higgs bosons decay into the same sign dilepton. 
The most stringent lower bound for the mass of doubly-charged Higgs bosons is about 400 GeV given at LHC~\cite{H++_LHC}. 
In our model, $\Delta^{\pm\pm}$ cannot decay into the same sign dilepton associated without any other particles, 
because they are the $\mathbb{Z}_2$ odd particles. 
In that case, the mass bound given at LHC cannot be applied to that in our model. 

First, we discuss the decay of $\Delta^{\pm\pm}$ in the scenario based on the lighter neutral component of $\Delta$ assumed to be DM. 
Basically, 
there are two decay modes of $\Delta^{\pm\pm}$: 
(1) the same sign dilepton decay through the Yukawa coupling $y_\Delta$ ($\Delta^{\pm\pm}\to E_4^\pm \ell^\pm$)\footnote{
The magnitudes of the branching fractions of $\Delta^{\pm\pm}$ in the same sign dilepton modes: $E_4^\pm e^\pm$, $ E_4^\pm \mu^\pm$ and $E_4^\pm \tau^\pm$ depend on the value of $y_\Delta^i$. 
In the following discussion for the collider phenomenology, we do not specify the flavor of $\ell^\pm$.  } 
with $\ell^\pm$ to be $e^\pm$, $\mu^\pm$ or $\tau^\pm$, 
and (2) W boson associated decay through the gauge coupling constant ($\Delta^{\pm\pm}\to \Delta^\pm W^{\pm*}$).
The decay branching fractions for (1) and (2) are determined by the magnitude of $y_\Delta$, $\Delta m$ and 
the mass of the exotic lepton $M$. 
The formulae for the decay rates for these channels are given in Appendix~B. 

In Fig.~\ref{decay}, the branching fraction of $\Delta^{\pm\pm}$ is
shown as a function of $M$. 
We take $m_{\Delta^{++}}=90.5$ GeV and $m_{\Delta^+}=78$ GeV which correspond to the case with $m_{\Delta^0}=63$ GeV and 
$\Delta m=15$ GeV. 
The Yukawa coupling $y_\Delta$ is taken to be 0.1 and 0.01.
It is seen that the main decay mode is changed from 
$\Delta^{++}\to E_4^+ \ell^+$ to $\Delta^{++}\to \Delta^+ W^{+*}$ when $M$ is getting larger values. 
For example, 50\% of $\mathcal{B}(\Delta^{++}\to E_4^+\ell^+)$ can be obtained in the case of  $M\simeq 89$ GeV (84 GeV)
and $y_\Delta=0.1$ (0.01).  


\begin{figure}[t]
\begin{center}
\includegraphics[width=100mm]{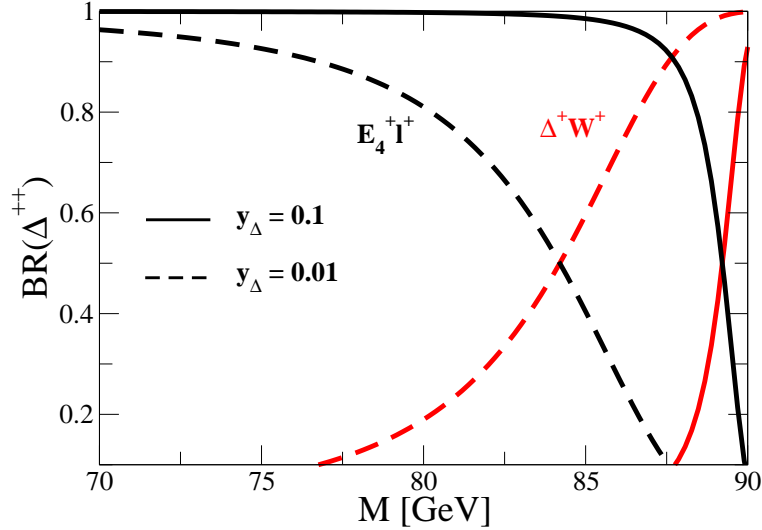}
\caption{The decay branching ratio of $\Delta^{\pm\pm}$ as a function of the exotic lepton mass $M$ 
for the case of $m_{\Delta^{++}}=90.5$ GeV, $m_{\Delta^{+}}=78$ GeV. 
The solid and dashed curves respectively show the cases with $y_\Delta=0.1$ and 0.01.}
\label{decay}
\end{center}
\end{figure}

In the following, we discuss the case where $\Delta^{\pm\pm}$ mainly decay into the same sign dilepton $(\Delta^{\pm\pm}\to \ell^\pm E_4^\pm)$. 
In that case, the exotic lepton mass $M$ should be between $m_{\Delta^{++}}$ and $m_{\Delta^+}$, otherwise 
there is no possible decay channel for $E_4^\pm$. 
As an benchmark scenario, we take the following mass spectrum and the coupling constant which is allowed from 
the electroweak precision data and also the LFV data: 
\begin{align}
&m_{\Delta^{++}}=90.5~\text{GeV},~~m_{\Delta^{+}}=78~\text{GeV},~~m_{\Delta^{0}}=63~\text{GeV},~~M=85~\text{GeV},\notag\\
&y_\Delta = 0.1. \label{set}
\end{align}
In this parameter set, the branching fraction of $\Delta^{\pm\pm}\to E_4^\pm\ell^\pm$ is about 99\%, and 
that of $\Delta^{\pm}\to W^\pm \Delta^0$ and $E_4^\pm \to \Delta^\pm \nu$ are 100\%, where 
$\Delta^0$ is $\Delta_R$ or $\Delta_I$. 
Thus, the decay process of $\Delta^{++}$ is expected as follows: 
\begin{align}
\Delta^{++}\to \ell^+ E_4^+\to \ell^+\Delta^+ \nu\to\ell^+W^+\Delta^0\nu \to \ell^+\ell^+\Delta^0\nu\nu, 
\end{align}
so that the final state contains the same sign dilepton and the missing energy. 

We then discuss the signal of $\Delta^{\pm\pm}$ at LHC in the parameter set given in Eq.~(\ref{set}). 
At LHC, $\Delta^{\pm\pm}$ can be mainly produced via the Drell-Yan processes: 
$q\bar{q}\to \gamma^*/Z^*\to \Delta^{++}\Delta^{--}$ and 
$q\bar{q}'\to W^{\pm*} \to \Delta^{\pm\pm}\Delta^{\mp}$. 
Thus, the signal events are expected to be\footnote{
The heavier $\Delta^0$; $e.g.,$ $\Delta_R$ can further decay into $Z^*$ and $\Delta_I$ which is corresponding to DM. }
\begin{align}
q\bar{q}&\to \Delta^{++}\Delta^{--}\to (\ell^+W^{+*}\nu \Delta^0)(\ell^-W^{-*}\nu\Delta^0)
\to \ell^+ \ell^+ \ell^-\ell^- E_T\hspace{-4.5mm}/\hspace{2mm},  \label{sig1}\\
q\bar{q}'&\to \Delta^{\pm\pm}\Delta^{\mp}\to (\ell^\pm W^{\pm*}\nu \Delta^0)(W^{\mp*} \Delta^0)\to
\ell^\pm \ell^\pm  \ell^\mp E_T\hspace{-4.5mm}/\hspace{2mm}. \label{sig2}
\end{align}
The cross sections of $\sigma(pp\to \Delta^{++}\Delta^{--})$, $\sigma(pp\to \Delta^{++}\Delta^{-})$ and 
$\sigma(pp\to \Delta^{--}\Delta^{+})$
are respectively evaluated as 566 fb, 889 fb and 494 fb with the collision energy to be 8 TeV by using 
{\tt CalcHEP}~\cite{CalcHEP} and {\tt CTEQ6L} parton distribution functions. 
The cross sections for the final states expressed in Eqs.~(\ref{sig1}) and (\ref{sig2}) 
are obtained as 63 fb and 154 fb, respectively. 

The same events can happen in HTM. 
The doubly-charged Higgs bosons $H^{\pm\pm}$ in HTM can decay into the same sign diboson $H^{\pm\pm}\to W^\pm W^\pm$ which 
is realized in the parameter regions where VEV of the triplet field $v_\Delta$ is larger than about $10^{-4}$ GeV, and $H^{\pm\pm}$ 
is the lightest of all the Higgs bosons from the triplet field. 
At the same time, singly-charged Higgs bosons $H^\pm$ in HTM can decay into $W^\pm Z$ in this case as long as the 
mass difference between $H^\pm$ and $H^\pm$ is not too large. 
Therefore, the same events as expressed in Eqs.~(\ref{sig1}) and (\ref{sig2}) can appear in the following way: 
\begin{align}
pp &\to H^{++}H^{--}\to W^+W^+W^-W^- \to \ell^+\ell^+\ell^-\ell^- E_T\hspace{-4.5mm}/\hspace{2mm}, \label{sig1_cascade}\\
pp &\to H^{\pm\pm}H^{\mp}\to W^\pm W^\pm W^\mp Z \to \ell^\pm \ell^\pm \ell^\mp  E_T\hspace{-4.5mm}/\hspace{2mm}, 
\end{align}
Furthermore, in the case where
$H^{\pm\pm}$ are the heaviest among the 
triplet like Higgs bosons such like 
our scenario, the cascade decay of $H^{\pm\pm}$ can be dominant; $i.e.,$ $H^{++}\to W^\pm H^\pm \to W^\pm W^\pm H^0$ ($H^0$ is a 
neutral component field of the triplet Higgs). 
In addition, $v_\Delta \lesssim 10^{-4}$ GeV, $H^0$ mainly decays into neutrinos. 
%
In this case, the final states of the signal event can also be 
\begin{align}
pp &\to H^{++}H^{--}\to (W^+W^+H^0)(W^-W^- H^0)\to \ell^+\ell^+ \ell^-\ell^- E_T\hspace{-4.5mm}/\hspace{2mm}, 
\label{sig1_diboson}\\
pp &\to H^{\pm\pm}H^{\mp}\to (W^\pm W^\pm H^0)(W^\mp H^0) \to \ell^\pm \ell^\pm \ell^\mp E_T\hspace{-4.5mm}/\hspace{2mm}. 
\end{align}

\begin{figure}[t]
\begin{center}
\includegraphics[width=120mm]{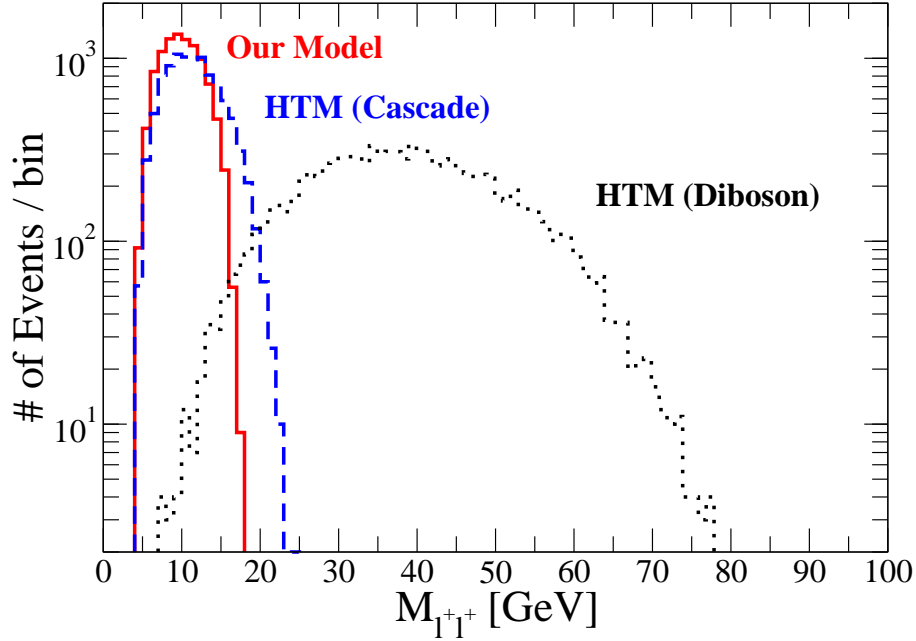}
\caption{Invariant mass distribution for the $\ell^+\ell^+$ system from 
the event $pp\to \ell^+\ell^+\ell^-\ell^-E_T\hspace{-4.5mm}/\hspace{2mm}$ 
in our model (red solid curve) 
and that in HTM with the diboson decay (black dotted curve) of $H^{\pm\pm}$ and 
the cascade decay of $H^{\pm\pm}$ (blue dashed curve). 
We take the mass of the doubly-charged scalar bosons to be 90.5 GeV and the collision energy to be 8 TeV. 
The solid, dashed and dotted curves are respectively shown the distributions for the dilepton decay 
in our model, cascade decay in the HTM and diboson decay in HTM. 
Number of events are assumed to be $10^{4}$ in each distribution. 
}
\label{Mee}
\end{center}
\end{figure}

The invariant mass distributions in the system of the same sign dilepton $M_{\ell^+\ell^+}$ can be useful to 
discriminate between our model and HTM.  
In Fig.~\ref{Mee}, the distribution for $M_{\ell^+\ell^+}$ in the $\ell^+\ell^+\ell^-\ell^- E_T\hspace{-4.5mm}/\hspace{2mm}$
system is shown in which the number of event is assumed to be $10^{4}$ and collision energy to be 8 TeV.
The mass of the doubly-charged scalar bosons in our model and also in HTM are taken to be 90.5 GeV. 
The red solid curve and the black dotted (blue dashed) curve are respectively represent 
the $M_{\ell^+\ell^+}$ distribution from our model and HTM with the diboson decay (cascade decay) of $H^{\pm\pm}$. 
It can be seen that the event number 
from the diboson decay of $H^{\pm\pm}$ is distributed in the wide region of $M_{\ell^+\ell^+}$, and 
it takes the maximum value at around the half of the mass of $H^{\pm\pm}$; $i.e., $ $M_{\ell^+\ell^+}\sim 45$ GeV. 
On the other hand, the other two shapes of the distributions look like each other in which the event number is distributed in the small region 
of $M_{\ell^+\ell^+}$. 
However, the cross sections for the final states are different between these two events. 
In our model, one of the two leptons with the same sign comes from the decay of $\Delta^{\pm\pm}$, and the other one comes from 
the leptonic decay of the W boson, so that the cross section is calculated as 
$\sigma(pp\to \Delta^{++}\Delta^{--})\times \mathcal{B}(W\to \ell\nu)^2$. 
On the other hand, in HTM with the cascade decay of $H^{\pm\pm}$, 
all the leptons in the final state are obtained from the leptonic decay of the W boson, 
so that the cross section is evaluated by $\sigma(pp\to H^{++}H^{--})\times \mathcal{B}(W\to \ell\nu)^4$.
The original cross sections of $\sigma(pp\to \Delta^{++}\Delta^{--})$ and $\sigma(pp\to H^{++}H^{--})$ 
are the same each other as long as we take the masses of $\Delta^{\pm\pm}$ and $H^{\pm\pm}$ to be the same. 
Therefore, the event number in HTM with the cascade decay scenario is smaller than that in our model by 
the factor of $\mathcal{B}(W\to \ell\nu)^2=1/9$. 

We then conclude that 
the four lepton events with the missing energy $\ell^+\ell^+\ell^-\ell^-E_T\hspace{-4.5mm}/\hspace{2mm}$
from our model and HTM with the diboson decay and the cascade decay of $H^{\pm\pm}$ may be able to be discriminated 
by using the $M_{\ell^+\ell^+}$ distribution and the number of event. 


\section{Conclusions}
We have constructed a two-loop radiative  seesaw model that provides neutrino
masses in a TeV scale theory. We have also studied DM properties, 
in which our model has fermionic ($N^c$) or bosonic DM ($\eta^0,\ \Delta^0$) candidate with the same mass scale, which is at around $m_h/2$, from the constraint of WMAP and the direct detection search in XENON100.
We have also discussed Higgs phenomenology at LHC, in which the neutral scalar boson from the triplet field 
is assumed to be DM. 
In that case, the mass of the doubly-charged scalar bosons $\Delta^{\pm\pm}$ is constrained to be smaller than about 
90.5 GeV by the electroweak precision data. 
In this scenario, $\Delta^{\pm\pm}$ can mainly decay into 
the same sign dilepton with the missing energy where the exotic leptons $E_4^\pm$ appear the intermediate state of 
the decay process of $\Delta^{\pm\pm}$. 
We then considered how we can discriminate the signature of $\Delta^{\pm\pm}$ in our model from 
that of doubly-charged Higgs boson $H^{\pm\pm}$ from HTM. 
We have found that the distribution of the same sign dilepton system and the magnitude of the 
cross section for $\ell^+\ell^+\ell^-\ell^-E_T\hspace{-4.5mm}/\hspace{2mm}$ can be useful to distinguish between 
our model and HTM.

\section*{Acknowledgments}
H.O. thanks to Prof. Eung-Jin Chun and Dr. Takashi Toma for fruitful discussion.
Y.K. thanks Korea Institute for Advanced Study for the travel support and local hospitality
during some parts of this work.
K.Y. was supported in part by the National Science Council of R.O.C. under Grant No. NSC-101-2811-M-008-014.

\begin{appendix}
\section{Gauge boson two point functions}

We list the analytic formulae for the 1PI diagram contributions to the gauge boson two point functions at the one loop level 
which are necessary to calculate the $S$ and $T$ parameters. 
We ignore the mixing between the $\mathbb{Z}_2$-even scalar bosons; $i.e.$, $\alpha=0$. 
In that case, new physics contributions to the gauge boson two point functions can be 
separated from the SM ones. 
The new physics contributions are calculated by 
\begin{align}
\Sigma_T^{WW}(p^2) &=
\frac{1}{16\pi^2}\frac{g^2}{4}\Big[16p^2B_3(p^2,M,M)\notag\\
&+4c_\beta^2B_5(p^2,m_{\Delta^{++}},m_{\Delta^+})
+4s_\beta^2B_5(p^2,m_{\Delta^{++}},m_{\eta^+})\notag\\
&+(\sqrt{2}c_\beta c_\gamma + s_\beta s_\gamma)^2B_5(p^2,m_{\Delta^{+}},m_{\Delta_R})
+(\sqrt{2}c_\beta c_\delta + s_\beta s_\delta)^2B_5(p^2,m_{\Delta^{+}},m_{\Delta_I})\notag\\
&+(c_\beta c_\gamma + \sqrt{2}s_\beta s_\gamma)^2B_5(p^2,m_{\eta^{+}},m_{\eta_R})
+(c_\beta c_\delta + \sqrt{2}s_\beta s_\delta)^2B_5(p^2,m_{\eta^{+}},m_{\eta_I})\notag\\
&+(-\sqrt{2}c_\beta s_\gamma + s_\beta c_\gamma)^2B_5(p^2,m_{\Delta^{+}},m_{\eta_R})
+(-\sqrt{2}c_\beta s_\delta + s_\beta c_\delta)^2B_5(p^2,m_{\Delta^{+}},m_{\eta_I})\notag\\
&+(-\sqrt{2}s_\beta c_\gamma + c_\beta s_\gamma)^2B_5(p^2,m_{\eta^{+}},m_{\Delta_R})
+(-\sqrt{2}s_\beta c_\delta + c_\beta s_\delta)^2B_5(p^2,m_{\eta^{+}},m_{\Delta_I})
\Big],
\end{align}
\begin{align}
\Sigma_T^{\gamma\gamma}(p^2) &= \frac{e^2}{16\pi^2}
\Big[8p^2B_3(p^2,M,M)\notag\\
&+4B_5(p^2,m_{\Delta^{++}},m_{\Delta^{++}})
+B_5(p^2,m_{\Delta^{+}},m_{\Delta^{+}})+B_5(p^2,m_{\eta^+},m_{\eta^+})\Big], 
\end{align}
\begin{align}
\Sigma_T^{Z\gamma}(p^2) &= \frac{eg_Z}{16\pi^2}\Big[
\Big(\frac{1}{2}-s_W^2\Big)8p^2B_3(p^2,M,M)\notag\\
&+2(1-2s_W^2)B_5(p^2,m_{\Delta^{++}},m_{\Delta^{++}})
+\frac{1}{2}(s_\beta^2-2s_W^2)B_5(p^2,m_{\Delta^{+}},m_{\Delta^{+}})\notag\\
&+\frac{1}{2}(c_\beta^2-2s_W^2)B_5(p^2,m_{\eta^+},m_{\eta^+})\Big], 
\end{align}
\begin{align}
\Sigma_T^{ZZ}(p^2) &= \frac{1}{16\pi^2}\frac{g_Z^2}{4}\Big[
32p^2\Big(\frac{1}{2}-s_W^2+s_W^4\Big)B_3(p^2,M,M)\notag\\
&+4(1-2s_W^2)^2B_5(p^2,m_{\Delta^{++}},m_{\Delta^{++}})
+(s_\beta^2-2s_W^2)^2B_5(p^2,m_{\Delta^{+}},m_{\Delta^{+}})\notag\\
&+(c_\beta^2-2s_W^2)^2B_5(p^2,m_{\eta^+},m_{\eta^+})
+2c_\beta^2 s_\beta^2B_5(p^2,m_{\Delta^+},m_{\eta^+})\notag\\
&+(2c_\gamma c_\delta +s_\gamma s_\delta)^2B_5(p^2,m_{\Delta_R},m_{\Delta_I})
+(c_\gamma c_\delta +2s_\gamma s_\delta)^2B_5(p^2,m_{\eta_R},m_{\eta_I})\notag\\
&+(-2c_\gamma s_\delta +s_\gamma c_\delta)^2B_5(p^2,m_{\Delta_R},m_{\eta_I})
+(-2s_\gamma c_\delta +c_\gamma s_\delta)^2B_5(p^2,m_{\eta_R},m_{\Delta_I})
\Big], 
\end{align}
where $c_\theta=\cos\theta$ and $s_\theta=\sin\theta$. 
In the above equations, 
$B_3(p^2,m_1,m_2)$ and $B_5(p^2,m_1,m_2)$ functions~\cite{hhkm} are respectively expressed in terms of the Passarino-Veltman functions~\cite{PV} by 
\begin{align}
B_3(p^2,m_1,m_2)&=-B_1(p^2,m_1,m_2)-B_{21}(p^2,m_1,m_2),\\
B_5(p^2,m_1,m_2)&=A(m_1)+A(m_2)-4B_{22}(p^2,m_1,m_2). 
\end{align}
The functions $\Sigma_{T}^{11}$, $\Sigma_{T}^{3Q}$ and $\Sigma_{T}^{33}$ are given in terms of 
above the gauge boson two point functions by 
\begin{align}
\Sigma_T^{11}&=\frac{1}{g^2}\Sigma_T^{WW},\quad
\Sigma_T^{3Q}&=\frac{1}{g^2}\left[\frac{c_W}{s_W}\Sigma_T^{Z\gamma}+\Sigma_T^{\gamma\gamma}\right],\quad
\Sigma_T^{33}=\frac{1}{g^2}\left[c_W^2\Sigma_T^{ZZ}+2s_Wc_W\Sigma_T^{Z\gamma}+s_W^2\Sigma_T^{\gamma\gamma}\right].
\end{align}

\section{Decay rates}

The decay rates for the doubly-charged scalar bosons $\Delta^{\pm\pm}$ are calculated as 
\begin{align}
\Gamma(\Delta^{\pm\pm}\to \ell_i^\pm E_4^\pm)&=\frac{m_{\Delta^{++}}}{16\pi}|y_\Delta^i|^2
\left(1-\frac{m_{\ell_i}^2}{m_{\Delta^{++}}^2}-\frac{M^2}{m_{\Delta^{++}}^2}\right)
\lambda^{1/2}\left(\frac{m_{\ell_i}^2}{m_{\Delta^{++}}^2},\frac{M^2}{m_{\Delta^{++}}^2}\right),\\
\Gamma(\Delta^{\pm\pm}\to \Delta^\pm W^{\pm})
&=\frac{g^2}{16\pi}\frac{m_{\Delta^{++}}^3}{m_W^2}\cos^2\beta\lambda^{3/2}
\left(\frac{m_{\Delta^+}^2}{m_{\Delta^{++}}^2},\frac{m_W^2}{m_{\Delta^{++}}^2}\right),\\
\Gamma(\Delta^{\pm\pm}\to \Delta^\pm W^{\pm *})&=\frac{9g^4m_{\Delta^{++}}}{128\pi^3}\cos^2\beta G\left(\frac{m_{\Delta^+}^2}{m_{\Delta^{++}}^2},\frac{m_W^2}{m_{\Delta^{++}}^2}\right), \\
\Gamma(\Delta^{\pm\pm}\to \eta^\pm W^{\pm})
&=\frac{g^2}{16\pi}\frac{m_{\Delta^{++}}^3}{m_W^2}\sin^2\beta\lambda^{3/2}
\left(\frac{m_{\eta^+}^2}{m_{\Delta^{++}}^2},\frac{m_W^2}{m_{\Delta^{++}}^2}\right),\\
\Gamma(\Delta^{\pm\pm}\to \eta^\pm W^{\pm *})&=\frac{9g^4m_{\Delta^{++}}}{128\pi^3}\sin^2\beta G\left(\frac{m_{\eta^+}^2}{m_{\Delta^{++}}^2},\frac{m_W^2}{m_{\Delta^{++}}^2}\right), 
\end{align}
where $\lambda(x,y)$ and $G(x,y)$ are the phase space functions which are given by 
 \begin{align}
\lambda(x,y) &= 1+x^2+y^2-2x-2y-2xy,\\
G(x,y)&=\frac{1}{12y}\Bigg\{2\left(-1+x\right)^3-9\left(-1+x^2\right)y+6\left(-1+x\right)y^2\notag\\
&+6\left(1+x-y\right)y\sqrt{-\lambda(x,y)}\left[\tan^{-1}\left(\frac{-1+x-y}{\sqrt{-\lambda(x,y)}}\right)+\tan^{-1}\left(\frac{-1+x+y}{\sqrt{-\lambda(x,y)}}\right)\right]\notag\\
&-3\left[1+\left(x-y\right)^2-2y\right]y\log x\Bigg\}. 
\end{align}

\section{Flavor symmetry}

\begin{table}[thbp]
\centering {\fontsize{10}{12}
\begin{tabular}{||c|c|c|c|c|c||c|c|c|c|c|}
\hline\hline ~~Particle~~ ~~ & ~~$L$~~ & ~~$e^c$~~ & ~~ $N^c$~~
  & $L'$  & $L'^c$  & ~~$\Delta$~~ & ~~$\Phi_i~~ $& ~~$\eta~~ $ & $\chi_1^{(i)} $ & $\chi_2^{(i)} $\\\hline
$T_7$  & $\bar3$ & $3$  & $3$  & $3$ & ${\bar 3}$
& $1_0$  & $1_{i}$  & $1_0$ & $\bar 3$ & $3$
\\\hline
\end{tabular}%
} \caption{The particle assignments in $T_7$ symmetry. Here $i$ runs 0-2.}
\label{t7}
\end{table}

Here we show an example how to realize our model for the lepton sector by using non-Abelian discrete symmetry.
The minimal extension is to introduce $T_7$ flavor symmetry \cite{Luhn:2007sy,Cao:2010mp, Cao:2011cp, Ishimori:2012sw}. 
Each of the field assignment is given in Table \ref{t7}, where the other assignments are same. 
The extended Lagrangian to Eq. (\ref{main-lag}) is modified as
\footnote{The charged-lepton sector has to be improved in order to forbid the universal Yukawa coupling of the SM-like Higgs boson, 
since it has been ruled out by the current Higgs boson search data at LHC. 
Straightforward ways to solve it is to change the flavor symmetry group and/or introduce some additional Higgs fields.}
\bea
{\cal L}_{N^c}&=& 
\sum_{i=1}^3\Phi^\dag_{i-1} \left(y_{\ell}^{a}+\omega^{2(i-1)}y_{\ell}^{b}+\omega^{i-1}y_{\ell}^{c}\right)e^c_i L_i
+ y_{\nu}\sum_{i=1}^3\eta^\dag N^c_i L'^c_i + y_{\Delta}\sum_{i=1}^3 \bar L^c_i i\tau_2\Delta L'_i  + M  \sum_{i=1}^3L'_i  L'^c_i 
\nn\\
&+&y_S^a \left(\chi^{(1)}_1 N^c_3N^c_3+\chi^{(2)}_1 N^c_1N^c_1+ \chi^{(3)}_1 N^c_2N^c_2\right)\notag\\
&+&
y_S^b \left(\chi^{(1)}_2 (N^c_2N^c_3 +N^c_3N^c_2)
+\chi^{(2)}_2 (N^c_3N^c_1+N^c_1N^c_3)+ 
\chi^{(3)}_2 (N^c_1N^c_2+N^c_2N^c_1)\right)+\mathrm{h.c.},
\eea
where $\omega\equiv e^{2i\pi/3}$. Here notice that all the terms except $\chi N^cN^c$ are diagonal. 
It suggests that the observed neutrino mass and lepton mixing can be obtained only through the $\chi N^cN^c$ term.
After the $B-L$ spontaneously breaking; $\langle \chi_1^{(i)}\rangle=v^{(i)}_1$ and $\langle \chi_2^{(i)}\rangle=v^{(i)}_2$, 
the right-handed neutrino mass matrix is given by
\be
m_{N^c}
=
y_S^a\left(%
\begin{array}{ccc}
v_1^{(2)} & 0 & 0 \\
0& v_1^{(3)} & 0 \\
0 & 0& v_1^{(1)}
\end{array}%
\right)
+
y_S^b\left(%
\begin{array}{ccc}
0 & v_2^{(3)}  & v_2^{(2)} \\
v_2^{(3)} & 0 & v_2^{(1)} \\
v_2^{(2)} & v_2^{(1)}& 0
\end{array}%
\right).
\ee
As a result, we can easily find the observed neutrino mass difference and their mixings by controlling each of the VEV.
In the Higgs potential, the modifications are as follows:
\be
\Phi^\dag\Phi\to\sum_{i=1}^3 \Phi^\dag_i\Phi_i,\quad \chi^\dag\chi\to \sum_{i=1}^2\sum_{j=1}^3 \chi^{\dag(j)}_i\chi_i^{(j)}.
\ee

\end{appendix}

\end{document}